\documentclass[sigconf,screen]{acmart}

\usepackage{graphicx}
\usepackage{xcolor}
\usepackage{multirow}
\usepackage[inline]{enumitem}
\usepackage{subcaption}
\usepackage{sistyle}
\usepackage[ruled]{algorithm2e}
\usepackage{booktabs}
\usepackage{caption}
\SIthousandsep{,}
\usepackage{makecell}
\usepackage[english]{babel}
\usepackage{algorithmic}
\usepackage{tikz}
\usepackage{wrapfig}
\usepackage{acronym}
\usepackage{CJKutf8}
\usepackage{hyperref}
\hypersetup{
    colorlinks,
    citecolor=blue
}
\usepackage{circledsteps}
\pgfkeys{/csteps/fill color=black}
\pgfkeys{/csteps/inner color=white}

\usepackage{colortbl}
\newcommand{\heading}[1]{\vspace*{.5mm}\noindent\textbf{#1.}}

\AtBeginDocument{%
  \providecommand\BibTeX{{%
    \normalfont B\kern-0.5em{\scshape i\kern-0.25em b}\kern-0.8em\TeX}}}

\definecolor{lightred}{rgb}{1, 0.8, 0.8}

\makeatletter
\g@addto@macro\normalsize{%
  \abovedisplayskip 3pt plus1pt %minus1pt%
  \belowdisplayskip 3pt plus1pt
  \abovedisplayshortskip  0pt plus1pt%
  \belowdisplayshortskip  0pt plus1pt% minus1pt%
}
\makeatother

\acrodef{CV}{computer vision}
\acrodef{IR}{information retrieval}
\acrodef{LLM}{large language model}
\acrodef{MDP}{Markov decision process}
\acrodef{NLP}{natural language processing}
\acrodef{NRM}{neural ranking model}
\acrodef{RL}{reinforcement learning}

\author{Hongru Song}
\orcid{0009-0000-8443-9499}
\author{Ruqing Zhang}
\authornote{Ruqing Zhang is the corresponding author.}
\orcid{0000-0003-4294-2541}
\affiliation{
 \institution{State Key Laboratory of AI Safety, Institute of Computing Technology, Chinese Academy of Sciences}
 \institution{University of Chinese Academy of Sciences}
 \city{Beijing}
 \country{China}
}
\email{{songhongru24s,zhangruqing}@ict.ac.cn}

\author{Jiafeng Guo}
\orcid{0000-0002-9509-8674}
\author{Xueqi Cheng}
\orcid{0000-0002-5201-8195}
\affiliation{
 \institution{State Key Laboratory of AI Safety, Institute of Computing Technology, Chinese Academy of Sciences}
 \institution{University of Chinese Academy of Sciences}
 \city{Beijing}
 \country{China}
}
\email{{guojiafeng,cxq}@ict.ac.cn}

\author{Maarten de Rijke}
\orcid{0000-0002-1086-0202}
\affiliation{
 \institution{University of Amsterdam}
 \city{Amsterdam}
 \country{The Netherlands}
}
\email{m.derijke@uva.nl}

\renewcommand{\shortauthors}{Song et al.}

\copyrightyear{2026}
\acmYear{2026}
\setcopyright{cc}
\setcctype{by}

\acmConference[CIKM '26]
{Proceedings of the 35th ACM International Conference on Information and Knowledge Management}
{November 07--11, 2026}
{Rome, Italy}

\acmBooktitle{Proceedings of the 35th ACM International Conference on Information and Knowledge Management (CIKM '26), November 07--11, 2026, Rome, Italy}

\acmDOI{10.1145/3799682.3840257}

\acmISBN{979-8-4007-2539-5/2026/11}



\begin{document}

\title[DeepRepro: State-Aware Subplanning for Paper-to-Code Reproduction in Evolving Repositories]{DeepRepro: State-Aware Subplanning for Paper-to-Code Reproduction in Evolving Repositories}

\begin{abstract}
Recent advances in agentic large language models (LLMs) have enabled increasingly autonomous software engineering workflows, yet automatic machine learning (ML) paper-to-code reproduction remains a challenging long-horizon problem. Unlike conventional code generation, this task requires constructing and maintaining a fully functional repository whose state continuously evolves during execution. 
Existing systems typically rely on static upfront planning followed by sequential file-level generation, which often leads to inconsistencies as dependencies, interfaces, and execution feedback change over time. 
We propose DeepRepro, a state-aware framework for paper-to-code reproduction based on execution-state-aware subplanning. DeepRepro dynamically transforms evolving repository states and runtime feedback into fine-grained implementation subplans, keeping planning aligned with execution throughout repository construction. The framework further incorporates repository-aware orchestration and a lightweight process-aware interface for transparent monitoring of long-horizon reproduction. 
Experiments on PaperBench Code-Dev show that DeepRepro consistently outperforms strong scientific and commercial code-agent baselines. 
Code and demo are available at \url{https://github.com/ruyisy/DeepRepro} and \url{https://youtu.be/ZvmtVX7Rleo}.

\end{abstract}

\maketitle

\vspace{-1.5mm}
\section{Introduction}

\vspace*{-0.5mm}\noindent%
Recent advances in agentic large language models (LLMs) are rapidly transforming code assistants from passive completion tools into autonomous software engineering agents capable of building full repositories from high-level instructions \cite{dong2025surveycodegenerationllmbased,ge2025surveyvibecodinglarge,qian2024chatdevcommunicativeagentssoftware,NEURIPS2024_5a7c9475}. 
A particularly challenging instantiation of this trend is \emph{scientific paper-to-code reproduction}, where an agent must reconstruct a complete, executable codebase from a research paper description \cite{seo2026paper2codeautomatingcodegeneration,liDeepCodeOpenAgentic2025a}. 
Unlike traditional code generation, this task requires not only understanding algorithms, but also assembling interdependent modules, resolving implicit implementation details, and producing reproducible experimental pipelines. 

\begin{figure*}[t]
\centering
\includegraphics[width=\textwidth]{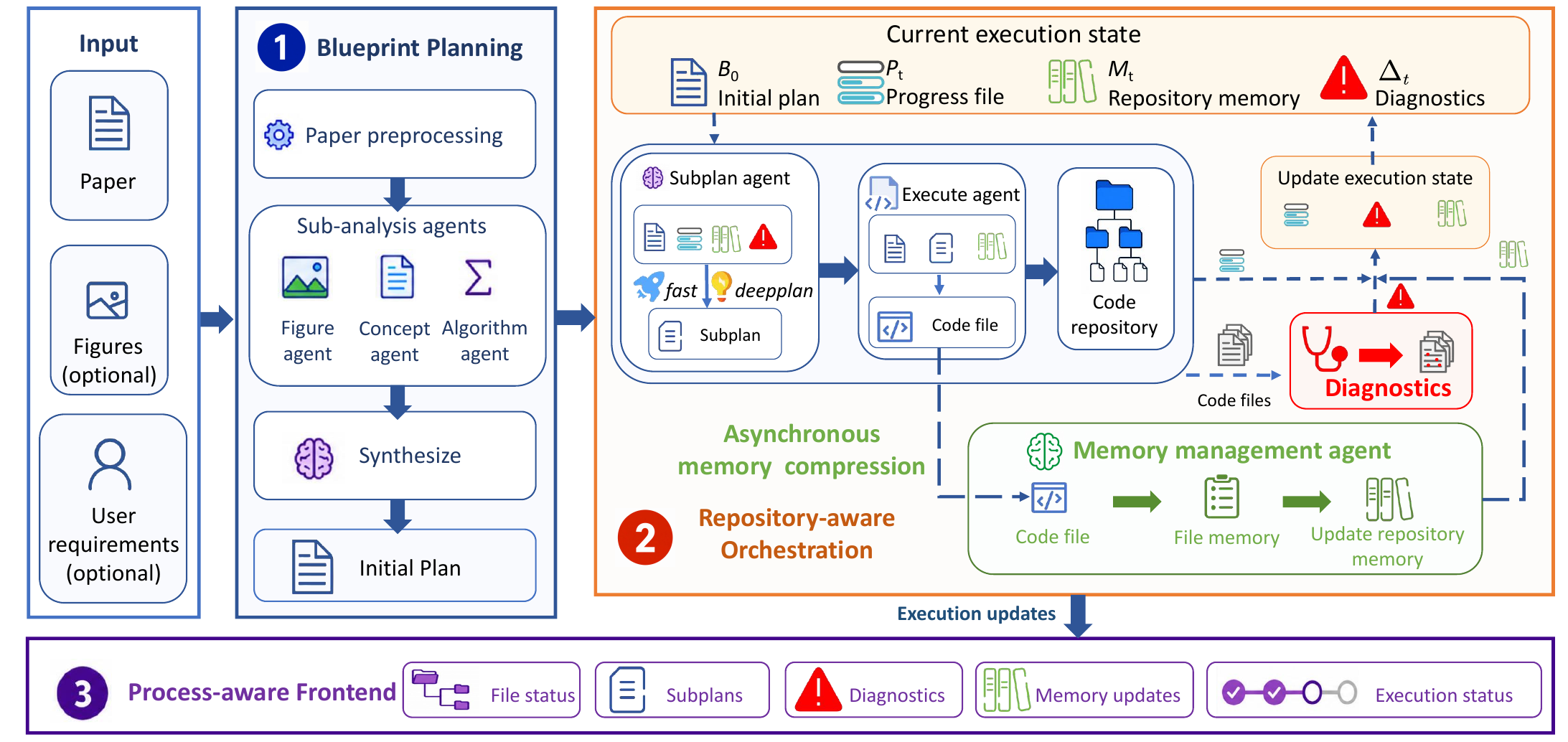}
\caption{Overview of the DeepRepro framework for paper-to-code reproduction. }
\label{fig:framework}
\vspace{-4mm}
\end{figure*}

Recent benchmarks such as PaperBench provide a structured evaluation setting with fine-grained rubrics for measuring reproduction quality \cite{starace2025paperbenchevaluatingaisability}. 
Despite progress in agentic coding systems, even strong iterative agents still fall far behind human experts. 
Recent specialized frameworks improve over general-purpose coding agents, yet their performance remains limited in long-horizon reproduction settings \cite{seo2026paper2codeautomatingcodegeneration,zhao2026autoreproduceautomaticaiexperiment}. 
A key limitation is that most existing systems rely heavily on an initial global plan, and then proceed with incremental repository construction.

However, in paper-to-code reproduction, the repository is a dynamic and evolving artifact. 
As code is generated, new interfaces emerge, dependencies become concrete, and earlier assumptions may become invalid. 
As a result, a static plan quickly becomes misaligned with the actual implementation state, leading to cascading inconsistencies and repeated repair cycles. 
In addition, since reproduction is long-running and resource-intensive, users also lack a clear view into intermediate progress, partial failures, and system-level diagnostics. 
This reveals a fundamental challenge: \emph{effective paper-to-code reproduction requires continual alignment between planning and an evolving codebase state, rather than a one-shot decomposition followed by static execution}.

To this end, we propose DeepRepro, a state-aware framework for paper-to-code reproduction based on iterative subplanning over evolving repositories. 
The key idea is to treat repository construction as an iterative process where planning is continuously conditioned on the current implementation state.
As shown in Figure~\ref{fig:framework}, DeepRepro consists of three tightly coupled components:
\begin{enumerate*}[label=\Circled{\arabic*}]
\item \textit{Blueprint planning} first distills a research paper into a repository-level skeleton, including core modules, execution flow, dependencies, and evaluation protocols. 
\item \textit{Repository-aware orchestration} then constructs the repository through a state-conditioned iterative loop. 
At each round, a subplanning module dynamically selects a set of tightly coupled files or repair targets based on the current repository state; an execution module implements them using tool-augmented generation; a memory module compresses completed artifacts into structured repository memory; and a bounded repair mechanism feeds execution diagnostics back into subsequent planning steps. 
This design supports both lightweight file-level execution (\textit{fast mode}) and deeper dependency-aware grouping and ordering (\textit{deepplan mode}). And
\item a \textit{process-aware frontend} exposes the intermediate reproduction process, including file status, evolving subplans, and runtime diagnostics, enabling transparent monitoring of long-horizon execution.  
\end{enumerate*}

We evaluate DeepRepro on PaperBench~\cite{starace2025paperbenchevaluatingaisability}. 
On the five-paper subset, DeepRepro achieves an average score of 84.2, improving over DeepCode, Cursor, and Codex by 1.3, 17.7, and 23.2 points, respectively; on the shared human-expert subset, it exceeds the reported human expert results by 5.4 points. 
Further analysis shows that deep subplanning brings a 2.15-point improvement.

\section{The DeepRepro Framework}
Given $\mathcal{D}=(p,\mathcal{F},u)$, where $p$ is the input paper, $\mathcal{F}$ denotes optional figures, and $u$ denotes optional user requirements, DeepRepro aims to synthesize a repository $\mathcal{R}^{*}=\arg\max_{\mathcal{R}}\mathrm{Score}(\mathcal{R}\mid\mathcal{D})$. 
As shown in Figure~\ref{fig:framework}, DeepRepro first constructs a blueprint-level plan $B_0$, and then performs repository construction over the evolving state.

\vspace{-2mm}
\subsection{Blueprint planning}
\vspace*{-0.5mm}
\textbf{Paper input and preprocessing.}
Blueprint planning starts by normalizing heterogeneous inputs. 
Given a local paper file or URL, optional user requirements, and auxiliary figures, DeepRepro creates a task-specific workspace, copies or downloads the paper, and converts it into a Markdown document for later agents. 
It then standardizes paths for planning and implementation artifacts, and chooses either full-document reading or section-based segmented reading according to the document size and preprocessing status. 
The output is a clean workspace and a preprocessed document representation $\tilde{\mathcal{D}}$ for initial planning.

\heading{Paper analysis and initial planning}
DeepRepro uses a decom\-pose-and-summarize pipeline to build the initial reproduction blueprint. 
Three analysis agents produce complementary evidence views:
\begin{equation}
Z_{\mathrm{fig}}=A_{\mathrm{fig}}(\mathcal{F}),\quad
Z_{\mathrm{con}}=A_{\mathrm{con}}(\tilde{\mathcal{D}}),\quad
Z_{\mathrm{alg}}=A_{\mathrm{alg}}(\tilde{\mathcal{D}}),
\end{equation}
where $A_{\mathrm{fig}}$ extracts visual evidence from figures, $A_{\mathrm{con}}$ summarizes the problem, method, datasets, metrics, and results, and $A_{\mathrm{alg}}$ extracts implementation-critical details such as algorithms, equations, model components, losses, training procedures, hyperparameters, and assumptions. 
The planning agent then synthesizes these views with user requirements into the initial plan:
\begin{equation}
B_0=A_{\mathrm{plan}}(Z_{\mathrm{fig}},Z_{\mathrm{con}},Z_{\mathrm{alg}},u).
\end{equation}
The resulting $B_0$ specifies the target repository structure, implementation order, core modules, interfaces, data flow, reproduction settings, environment requirements, and validation procedures. 
Following DeepCode \cite{liDeepCodeOpenAgentic2025a}, DeepRepro can also retrieve repositories associated with important cited papers as optional auxiliary references. 
Once planning finishes, $B_0$ serves as the global specification for the generation loop.

\vspace{-2mm}
\subsection{Repository-aware orchestration}

\vspace{-0.5mm}
Given the initial plan $B_0$, DeepRepro initializes the project file tree and progress file, and then enters an agentic execution loop. 
Each execution round corresponds to one local implementation step: the system selects a small set of related files or repair targets, implements them, updates repository memory, and feeds diagnostics into the next round.
At round $t$, the execution state is
\begin{equation}
S_t=(B_0,\mathcal{P}_t,\mathcal{M}_t,\Delta_t),
\end{equation}
where $\mathcal{P}_t$ records completed and remaining files, $\mathcal{M}_t$ is the repository memory, and $\Delta_t$ stores diagnostics and repair signals.

\heading{Subplanning}
During execution, the repository state evolves as files are completed, memory is accumulated, and diagnostics are collected. 
The subplan agent $A_{\mathrm{sub}}$ uses these signals to compute the next subplan:
\begin{equation}
B_t=A_{\mathrm{sub}}(B_0,\mathcal{P}_t,\mathcal{M}_t,\Delta_t).
\end{equation}
The subplan $B_t$ selects a target set $G_t$, which may contain new files or files to repair. 
If $G_t=\emptyset$, the agentic loop stops and DeepRepro finalizes the repository and reproduction report.
Otherwise, the selected files are implemented in the current round. 
The same interface supports two modes: in \textit{fast} mode, $B_t$ only specifies the target file set $G_t$ for efficient generation; in \textit{deepplan} mode, $B_t=(G_t,\rho_t,\pi_t,\mathcal{I}_t,\mathcal{N}_t,\mathcal{Q}_t)$ additionally includes the grouping rationale $\rho_t$, execution order $\pi_t$, per-file instructions $\mathcal{I}_t$, integration notes $\mathcal{N}_t$, and acceptance checks $\mathcal{Q}_t$.

\heading{Execution and memory update}
Given $B_t$, the execution agent $A_{\mathrm{exec}}$ implements each target file $c_{t,i}\in G_t$ with tool calls. 
For each file, it uses the file-level context $X_{t,i}=(B_0,B_t,\mathcal{M}_t,c_{t,i})$, where $B_0$ provides the global plan, $B_t$ provides the current subplan, $\mathcal{M}_t$ provides repository memory, and $c_{t,i}$ identifies the target file. 
The execute agent generates the file content as $\hat{c}_{t,i}=A_{\mathrm{exec}}(X_{t,i})$ and writes it to the target file $c_{t,i}$. 
After a successful write, DeepRepro records the file in $\mathcal{P}_{t+1}$ and launches the memory agent $A_{\mathrm{mem}}$ for asynchronous compression:
\begin{equation}
m_{t,i}=A_{\mathrm{mem}}(\hat{c}_{t,i}),\qquad 
\mathcal{M}_{t+1}=\mathcal{M}_t\cup\{m_{t,i}\}_{c_{t,i}\in G_t}.
\end{equation}
Each memory entry summarizes the file purpose, interfaces, dependencies, and implementation notes. 
Before the next round, pending memory updates are synchronized so that the next subplan observes the latest repository memory.

\heading{Diagnostics and repair}
After each round, DeepRepro collects diagnostics for the generated or updated files in $G_t$ as $\Delta_{t+1}=\Delta_t\cup A_{\mathrm{diag}}(G_t)$, including tool errors, invalid writes, syntax errors, dependency issues, and layout problems.
These diagnostics serve as repair signals for later subplans.
In \textit{deepplan} mode, a bounded quality gate further identifies blocking issues and turns them into localized repair targets for later subplanning and execution.

\vspace{-2mm}
\subsection{Process-aware frontend}
DeepRepro provides a process-aware frontend for monitoring the long-horizon generation process. 
As shown in Figure~\ref{fig:framework}, it receives execution updates from the repository-aware orchestration loop and exposes file status, subplans, diagnostics, memory updates, and execution status, allowing users to inspect progress and identify potential issues during reproduction.

\begin{figure*}[!t]
\centering
\includegraphics[width=\textwidth]{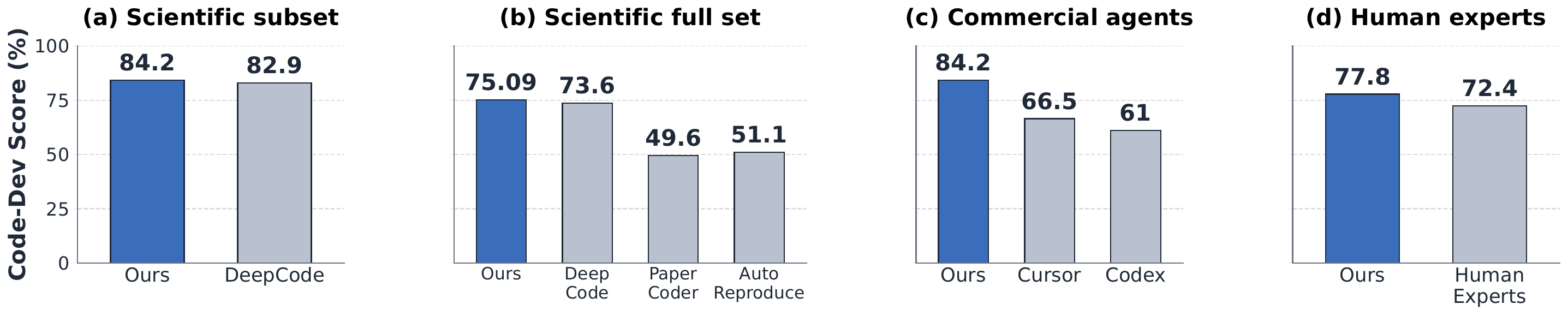}
\Description{Four bar charts comparing DeepRepro with scientific code agents, full-set scientific baselines, commercial code assistants, and human experts on PaperBench Code-Dev.}
\vspace*{-3mm}
\caption{Comparative Code-Dev results: (a) scientific-agent subset comparison, (b) full-set scientific-agent comparison, (c) commercial-agent comparison, and (d) human-expert comparison. DeepRepro is shown in blue and all baselines are shown in gray. Panel (b) includes full-set baselines from prior work and should not be interpreted as a strict comparison.}
\label{fig:comparative-experiments}
\vspace{-3mm}
\end{figure*}

\section{Experimental Setup}

\vspace*{-1mm}\noindent%
\textbf{Dataset and evaluation.}
We evaluate DeepRepro on PaperBench Code-Dev~\cite{starace2025paperbenchevaluatingaisability}, a code-development subset of PaperBench built from 20 ICML 2024 Spotlight and Oral papers. 
Following the official Code-Dev evaluation, we skip reproduction execution and grade only the Code Development requirements in the hierarchical rubrics, \textbf{with final scores reported as percentages}. 
We report full-set scores and fixed subset averages when needed; the five-paper subset includes FRE, RICE, PINN, BaM, and Mech-U.
For cost control, our main LLM-judge experiments use GPT-4.1-mini; for the human-expert comparison, we additionally use o3-mini for a fairer comparison with the PaperBench judging setup.

\heading{Baselines}
We compare DeepRepro with three groups of baselines:
\begin{enumerate*}[label=(\roman*)]
\item \textit{Scientific code agents}. DeepCode is our main scientific baseline \cite{liDeepCodeOpenAgentic2025a}, and we also compare with PaperCoder \cite{seo2026paper2codeautomatingcodegeneration} and AutoReproduce \cite{zhao2026autoreproduceautomaticaiexperiment};
\item \textit{Commercial code assistants}. We compare with Cursor and Codex, two general-purpose coding agents widely used in practice \cite{anysphere2025cursor,openai2025codex};
\item \textit{Human experts}. We use the PaperBench human baseline~\cite{starace2025paperbenchevaluatingaisability}, where eight ML PhD students or graduates worked part-time on three papers (FRE, All-in-One, and Stay-Topic). We report the best-of-three attempts as the expert-level reference.

\end{enumerate*}

\heading{Implementation details}
Unless otherwise specified, DeepRepro uses GPT-5.4 with high reasoning effort as the backbone model \cite{openai2026models}. In model-ablation experiments, the blueprint-planning stage is fixed, and only the subplan and execute models vary. All methods run without extra paper figures or user requirements, and agents do not consult the target paper's official source code. Following prior scientific code agent evaluations, we report both subset and full-set comparisons; some full-set numbers are reused from prior work because of evaluation cost, and we treat them as reference points rather than strict controlled claims.

\heading{Model variants}
The suffix \textit{+ref} means that external code references are enabled; \textit{deep\-plan-5.4} uses GPT-5.4 for both subplanning and execution; \textit{deepplan-5.4+DS} uses GPT-5.4 for subplanning and DeepSeek-V4-Flash \cite{deepseekai2026v4release} for execution; and \textit{fast-5.4}/\textit{fast-DS} use \textit{fast} mode with GPT-5.4/Deep\-Seek-V4-Flash execution.

\vspace{-2mm}
\section{Experimental Results}
\vspace{-.5mm}
\textbf{Overall performance.}
Table~\ref{tab:overall-performance} reports DeepRepro under four execution settings. We find:
\begin{enumerate*}[label=(\roman*)]
\item external code references have only marginal impact on this subset;
\item the \textit{deepplan} mode consistently outperforms \textit{fast}, improving the average score from 82.07 to 84.22.
\end{enumerate*}

\begin{table}[t]
\centering
\caption{Overall performance of DeepRepro under different execution modes on the five-paper subset. Highest scores in each column are boldfaced.}
\label{tab:overall-performance}
\setlength{\tabcolsep}{3.2pt}
\begin{tabular}{l cccccc}
\toprule
\textbf{Mode} & \textbf{FRE} & \textbf{RICE} & \textbf{PINN} & \textbf{BaM} & \textbf{Mech-U} & \textbf{Avg.} \\
\midrule
\textit{deepplan}+ref & 74.14 & \textbf{73.72} & \textbf{94.77} & \textbf{90.40} & 88.36 & \textbf{84.28} \\
\textit{deepplan} & \textbf{74.29} & 73.49 & 94.50 & 90.23 & \textbf{88.61} & 84.22 \\
\textit{fast}+ref & 70.15 & 71.86 & 94.05 & 87.18 & 87.32 & 82.11 \\
\textit{fast} & 69.46 & 72.51 & 93.95 & 87.20 & 87.23 & 82.07 \\
\bottomrule
\end{tabular}
%\vspace{-2mm}
\end{table}

\begin{table}[t]
\centering
\caption{Impact of subplanning and execution models. Highest scores in each column are boldfaced.}
\label{tab:llm-performance}
\setlength{\tabcolsep}{2.5pt}
\begin{tabular}{l cccccc}
\toprule
\textbf{Variant} & \textbf{FRE} & \textbf{RICE} & \textbf{PINN} & \textbf{BaM} & \textbf{Mech-U} & \textbf{Avg.} \\
\midrule
\textit{deepplan}-5.4 & \textbf{74.29} & \textbf{73.49} & \textbf{94.50} & \textbf{90.23} & \textbf{88.61} & \textbf{84.22} \\
\textit{deepplan}-5.4+DS & 67.18 & 72.90 & 92.92 & 86.10 & 86.87 & 81.15 \\
\textit{fast}-5.4 & 69.46 & 72.51 & 93.95 & 87.20 & 87.23 & 82.07 \\
\textit{fast}-DS & 64.24 & 68.93 & 89.31 & 82.74 & 82.60 & 77.56 \\
\bottomrule
\end{tabular}
\end{table}

\heading{Effect of subplanning and model choices} 
Table~\ref{tab:llm-performance} further analyzes how subplanning interacts with different execution models. 
We find that \textit{deepplan} consistently improves performance across both strong and weaker executors, and the gain becomes more pronounced when the execution model is weaker. This indicates that subplanning plays a complementary role by compensating for execution-level limitations and improving global coordination over long-horizon tasks.

\heading{Full-set evaluation}
Table~\ref{tab:full-set-scores} reports results on all 20 Code-Dev papers. DeepRepro achieves consistent performance across diverse tasks, with an overall average score of 75.09, further demonstrating robustness beyond the controlled five-paper setting.

\heading{Comparative experiments}
Figure~\ref{fig:comparative-experiments} summarizes the comparison results:
(i) \emph{Scientific code agents}. 
DeepRepro outperforms DeepCode on the controlled five-paper subset (Fig.~\ref{fig:comparative-experiments}(a)) and shows the same positive trend in the 20-paper setting (Fig.~\ref{fig:comparative-experiments} (b)), indicating robust gains across evaluation scales;
(ii) \emph{Commercial code assistants}. As shown in Fig.~\ref{fig:comparative-experiments} (c), DeepRepro surpasses Cursor and Codex under identical evaluation protocols on the five-paper subset;
(iii) \emph{Human experts}. Fig.~\ref{fig:comparative-experiments} (d) shows that DeepRepro exceeds the PaperBench Best@3 human baseline on the shared subset.

\heading{Runtime and cost analysis}
We further report approximate runtime and cost for practical reference. These numbers are not strictly controlled but reflect real deployment behavior. 
In \textit{fast} mode DeepRepro incurs similar costs to DeepCode (about US\$10$\pm$1 per paper), while the \textit{deepplan} mode increases costs moderately to around US\$12.5$\pm$1. In terms of runtime, DeepRepro significantly improves efficiency: \textit{fast} typically completes within one hour, while \textit{deepplan} finishes in approximately 1.5 hours, compared to nearly two hours for DeepCode. Overall, \textit{fast} is suitable for quick reproduction, while \textit{deepplan} is preferred for higher-fidelity results.

\begin{table}[t]
\centering
\caption{Full-set DeepRepro scores on 20 PaperBench papers.}
\label{tab:full-set-scores}
\setlength{\tabcolsep}{2.6pt}
\begin{tabular}{lrlrlr}
\toprule
\textbf{Paper} & \textbf{Score} & \textbf{Paper} & \textbf{Score} & \textbf{Paper} & \textbf{Score} \\
\midrule
FRE & 74.29 & RICE & 73.49 & PINN & 94.50 \\
Will-For. & 74.09 & BaM & 90.23 & All-in-One & 76.62 \\
Adapt. Prune & 56.60 & LBCS & 83.39 & Mech-U & 88.61 \\
TT Adapt. & 65.13 & Sample Spec. & 86.70 & Bridge Gap & 61.55 \\
Stay-Topic & 77.80 & Stochastic & 81.25 & LCA-Line & 78.37 \\
Seq. Neural & 87.44 & SAPG & 60.22 & FTRL & 41.91 \\
%\cline{5-6}
Robust-CLIP & 82.96 & BBox & 66.59 & \cellcolor[gray]{0.9}Avg. & \cellcolor[gray]{0.9}\textbf{75.09} \\
\bottomrule
\end{tabular}
\vspace{-1mm}
\end{table}

\vspace{-1mm}
\section{Related Work}

\vspace{-.5mm}\noindent\textbf{General coding agents.}
LLM-based coding agents have evolved from code completion to autonomous software engineering \cite{dong2025surveycodegenerationllmbased,ge2025surveyvibecodinglarge,guo2025comprehensivesurveybenchmarkssolutions}. 
Prior work explores role-based collaboration and workflow decomposition \cite{qian2024chatdevcommunicativeagentssoftware,hong2024metagptmetaprogrammingmultiagent}, repository-level editing and testing \cite{NEURIPS2024_5a7c9475}, and issue resolution through search, localization, task graphs, and iterative validation \cite{10.1145/3650212.3680384,chen2024coderissueresolvingmultiagent}. 
These capabilities have also been adopted by IDE- and terminal-based coding agents \cite{anysphere2025cursor,trae2025trae,windsurf2025windsurf,github2025copilot,cline2025cline,anthropic2025claudecode,openai2025codex,google2025geminicli}. 
Scientific reproduction is more demanding than general coding because agents must translate research contributions and experimental methodology into executable implementations \cite{seo2026paper2codeautomatingcodegeneration}. 

\noindent \textbf{Scientific coding agents.}
Scientific coding agents generate executable research artifacts from papers or experimental specifications. 
Paper-to-code systems have been evaluated through model-based assessment, human evaluation, and comparison with author repositories, while PaperBench standardizes evaluation using 20 ICML 2024 papers and hierarchical reproduction rubrics \cite{seo2026paper2codeautomatingcodegeneration,starace2025paperbenchevaluatingaisability}. 
Recent approaches include analysis--planning--generation pipelines \cite{seo2026paper2codeautomatingcodegeneration}, paper-lineage-based reproduction \cite{zhao2026autoreproduceautomaticaiexperiment}, autonomous research loops \cite{lu2024aiscientistfullyautomated,autoresearch,NEURIPS2025_0d904d30}, and planning with memory and verification \cite{liDeepCodeOpenAgentic2025a}.

\vspace{-1mm}
\section{Conclusion}
\vspace{-.5mm}
We have presented DeepRepro, an automatic framework for ML paper-to-code reproduction that treats repository construction as a state-aware, iterative process. 
DeepRepro combines blueprint planning with execution-state-aware round-level subplanning over evolving repositories, and integrates asynchronous memory compression, bounded repair, and a process-aware interface for long-horizon execution and monitoring. 
Experiments on PaperBench Code-Dev demonstrate that DeepRepro consistently outperforms strong scientific and commercial code-agent baselines.

\vspace*{-1mm}
\section*{Acknowledgements}
\vspace*{-.5mm}\noindent%
This work was funded by the Strategic Priority Research Program of the CAS under Grants No. XDB0680102, the National Natural Science Foundation of China (NSFC) under Grants No. 62472408, U25B2076 and 62441229, the National Key Research and Development Program of China under Grants No. 2023YFA1011602. This research was also (partially) supported by the Dutch Research Council (NWO), under project numbers 024.004.022, NWA.1389.20.183, and KICH3.LTP.20.006, and the European Union under grant agreement No. 101201510 (UNITE). 

All content represents the opinion of the authors, which is not necessarily shared or endorsed by their respective employers and/or sponsors.

\clearpage
\section*{GenAI Usage Disclosure}
We used generative AI tools to assist with grammar checking and language polishing during manuscript preparation. 

\bibliographystyle{ACM-Reference-Format}
\balance
\bibliography{references}

@String{Computing = "Computing" }

@misc{ge2025surveyvibecodinglarge,
      title={A Survey of Vibe Coding with Large Language Models}, 
      author={Yuyao Ge and Lingrui Mei and Zenghao Duan and Tianhao Li and Yujia Zheng and Yiwei Wang and Lexin Wang and Jiayu Yao and Tianyu Liu and Yujun Cai and Baolong Bi and Fangda Guo and Jiafeng Guo and Shenghua Liu and Xueqi Cheng},
      year={2025},
      eprint={2510.12399},
      archivePrefix={arXiv},
      primaryClass={cs.AI},
      url={https://arxiv.org/abs/2510.12399}, 
}

@misc{dong2025surveycodegenerationllmbased,
      title={A Survey on Code Generation with LLM-based Agents}, 
      author={Yihong Dong and Xue Jiang and Jiaru Qian and Tian Wang and Kechi Zhang and Zhi Jin and Ge Li},
      year={2025},
      eprint={2508.00083},
      archivePrefix={arXiv},
      primaryClass={cs.SE},
      url={https://arxiv.org/abs/2508.00083}, 
}

@misc{guo2025comprehensivesurveybenchmarkssolutions,
      title={A Comprehensive Survey on Benchmarks and Solutions in Software Engineering of LLM-Empowered Agentic System}, 
      author={Jiale Guo and Suizhi Huang and Mei Li and Dong Huang and Xingsheng Chen and Regina Zhang and Zhijiang Guo and Han Yu and Siu-Ming Yiu and Pietro Lio and Kwok-Yan Lam},
      year={2025},
      eprint={2510.09721},
      archivePrefix={arXiv},
      primaryClass={cs.SE},
      url={https://arxiv.org/abs/2510.09721}, 
}

@misc{qian2024chatdevcommunicativeagentssoftware,
      title={ChatDev: Communicative Agents for Software Development}, 
      author={Chen Qian and Wei Liu and Hongzhang Liu and Nuo Chen and Yufan Dang and Jiahao Li and Cheng Yang and Weize Chen and Yusheng Su and Xin Cong and Juyuan Xu and Dahai Li and Zhiyuan Liu and Maosong Sun},
      year={2024},
      eprint={2307.07924},
      archivePrefix={arXiv},
      primaryClass={cs.SE},
      url={https://arxiv.org/abs/2307.07924}, 
}

@misc{hong2024metagptmetaprogrammingmultiagent,
      title={MetaGPT: Meta Programming for A Multi-Agent Collaborative Framework}, 
      author={Sirui Hong and Mingchen Zhuge and Jiaqi Chen and Xiawu Zheng and Yuheng Cheng and Ceyao Zhang and Jinlin Wang and Zili Wang and Steven Ka Shing Yau and Zijuan Lin and Liyang Zhou and Chenyu Ran and Lingfeng Xiao and Chenglin Wu and Jürgen Schmidhuber},
      year={2024},
      eprint={2308.00352},
      archivePrefix={arXiv},
      primaryClass={cs.AI},
      url={https://arxiv.org/abs/2308.00352}, 
}

@inproceedings{NEURIPS2024_5a7c9475,
  title = {{{SWE-agent}}: {{Agent-computer}} Interfaces Enable Automated Software Engineering},
  booktitle = {Advances in Neural Information Processing Systems},
  author = {Yang, John and Jimenez, Carlos and Wettig, Alexander and Lieret, Kilian and Yao, Shunyu and Narasimhan, Karthik and Press, Ofir},
  editor = {Globerson, A. and Mackey, L. and Belgrave, D. and Fan, A. and Paquet, U. and Tomczak, J. and Zhang, C.},
  date = {2024},
  volume = {37},
  pages = {50528--50652},
  publisher = {Curran Associates, Inc.},
  doi = {10.52202/079017-1601},
  url = {https://proceedings.neurips.cc/paper_files/paper/2024/file/5a7c947568c1b1328ccc5230172e1e7c-Paper-Conference.pdf}
}

@inproceedings{10.1145/3650212.3680384,
author = {Zhang, Yuntong and Ruan, Haifeng and Fan, Zhiyu and Roychoudhury, Abhik},
title = {AutoCodeRover: Autonomous Program Improvement},
year = {2024},
isbn = {9798400706127},
publisher = {Association for Computing Machinery},
address = {New York, NY, USA},
url = {https://doi.org/10.1145/3650212.3680384},
doi = {10.1145/3650212.3680384},
booktitle = {Proceedings of the 33rd ACM SIGSOFT International Symposium on Software Testing and Analysis},
pages = {1592–1604},
numpages = {13},
location = {Vienna, Austria},
series = {ISSTA 2024}
}

@misc{chen2024coderissueresolvingmultiagent,
      title={CodeR: Issue Resolving with Multi-Agent and Task Graphs}, 
      author={Dong Chen and Shaoxin Lin and Muhan Zeng and Daoguang Zan and Jian-Gang Wang and Anton Cheshkov and Jun Sun and Hao Yu and Guoliang Dong and Artem Aliev and Jie Wang and Xiao Cheng and Guangtai Liang and Yuchi Ma and Pan Bian and Tao Xie and Qianxiang Wang},
      year={2024},
      eprint={2406.01304},
      archivePrefix={arXiv},
      primaryClass={cs.CL},
      url={https://arxiv.org/abs/2406.01304}, 
}

@misc{anysphere2025cursor,
  author       = {{Anysphere}},
  title        = {Cursor: The Best Way to Code with AI},
  howpublished = {\url{https://cursor.com}},
  year         = {2025},
  note         = {Accessed: 2026-05-19}
}

@misc{windsurf2025windsurf,
  author       = {{Windsurf}},
  title        = {Windsurf: The Best AI for Coding},
  howpublished = {\url{https://windsurf.com}},
  year         = {2025},
  note         = {Accessed: 2026-05-19}
}

@misc{trae2025trae,
  author       = {{TRAE}},
  title        = {TRAE: Collaborate with Intelligence},
  howpublished = {\url{https://www.trae.ai}},
  year         = {2025},
  note         = {Accessed: 2026-05-19}
}

@misc{anthropic2025claudecode,
  author       = {{Anthropic}},
  title        = {Claude Code: AI Coding Agent, Terminal, IDE},
  howpublished = {\url{https://claude.com/product/claude-code}},
  year         = {2025},
  note         = {Accessed: 2026-05-19}
}

@misc{openai2025codex,
  author       = {{OpenAI}},
  title        = {Codex CLI: A Coding Agent from OpenAI},
  howpublished = {\url{https://github.com/openai/codex}},
  year         = {2025},
  note         = {Accessed: 2026-05-19}
}

@misc{google2025geminicli,
  author       = {{Google}},
  title        = {Gemini CLI: An Open-Source AI Agent},
  howpublished = {\url{https://github.com/google-gemini/gemini-cli}},
  year         = {2025},
  note         = {Accessed: 2026-05-19}
}

@misc{github2025copilot,
  author       = {{GitHub}},
  title        = {GitHub Copilot: The AI Assistant That Builds with You},
  howpublished = {\url{https://github.com/copilot}},
  year         = {2025},
  note         = {Accessed: 2026-05-19}
}

@misc{cline2025cline,
  author       = {{Cline}},
  title        = {Cline: AI Coding, Open Source and Uncompromised},
  howpublished = {\url{https://cline.bot}},
  year         = {2025},
  note         = {Accessed: 2026-05-19}
}

@misc  {liDeepCodeOpenAgentic2025a,
  title = {{{DeepCode}}: {{Open Agentic Coding}}},
  shorttitle = {{{DeepCode}}},
  author = {Li, Zongwei and Li, Zhonghang and Guo, Zirui and Ren, Xubin and Huang, Chao},
  date = {2025-12-08},
  eprint = {2512.07921},
  eprinttype = {arXiv},
  eprintclass = {cs},
  doi = {10.48550/arXiv.2512.07921},
  url = {http://arxiv.org/abs/2512.07921},
  pubstate = {prepublished}
}

@misc{seo2026paper2codeautomatingcodegeneration,
      title={Paper2Code: Automating Code Generation from Scientific Papers in Machine Learning}, 
      author={Minju Seo and Jinheon Baek and Seongyun Lee and Sung Ju Hwang},
      year={2026},
      eprint={2504.17192},
      archivePrefix={arXiv},
      primaryClass={cs.CL},
      url={https://arxiv.org/abs/2504.17192}, 
}

@misc{zhao2026autoreproduceautomaticaiexperiment,
      title={AutoReproduce: Automatic AI Experiment Reproduction with Paper Lineage}, 
      author={Xuanle Zhao and Zilin Sang and Yuxuan Li and Qi Shi and Weilun Zhao and Shuo Wang and Duzhen Zhang and Xu Han and Zhiyuan Liu and Maosong Sun},
      year={2026},
      eprint={2505.20662},
      archivePrefix={arXiv},
      primaryClass={cs.AI},
      url={https://arxiv.org/abs/2505.20662}, 
}

@misc{starace2025paperbenchevaluatingaisability,
      title={PaperBench: Evaluating AI's Ability to Replicate AI Research}, 
      author={Giulio Starace and Oliver Jaffe and Dane Sherburn and James Aung and Jun Shern Chan and Leon Maksin and Rachel Dias and Evan Mays and Benjamin Kinsella and Wyatt Thompson and Johannes Heidecke and Amelia Glaese and Tejal Patwardhan},
      year={2025},
      eprint={2504.01848},
      archivePrefix={arXiv},
      primaryClass={cs.AI},
      url={https://arxiv.org/abs/2504.01848}, 
}

@misc{lu2024aiscientistfullyautomated,
      title={The AI Scientist: Towards Fully Automated Open-Ended Scientific Discovery}, 
      author={Chris Lu and Cong Lu and Robert Tjarko Lange and Jakob Foerster and Jeff Clune and David Ha},
      year={2024},
      eprint={2408.06292},
      archivePrefix={arXiv},
      primaryClass={cs.AI},
      url={https://arxiv.org/abs/2408.06292}, 
}

@misc{autoresearch,
  title        = {AutoResearch: AI Agents Running Research on Nanochat},
  author       = {Karpathy, Andrej},
  howpublished = {\url{https://github.com/karpathy/autoresearch}},
  year         = {2026},
  note         = {Accessed: 2026-05-19}
}

@inproceedings{NEURIPS2025_0d904d30,
  title = {{{AI-researcher}}: {{Autonomous}} Scientific Innovation},
  booktitle = {Advances in Neural Information Processing Systems},
  author = {Tang, Jiabin and Xia, Lianghao and Li, Zhonghang and Huang, Chao},
  editor = {Belgrave, D. and Zhang, C. and Lin, H. and Pascanu, R. and Koniusz, P. and Ghassemi, M. and Chen, N.},
  date = {2025},
  volume = {38},
  pages = {9481--9520},
  publisher = {Curran Associates, Inc.},
  url = {https://proceedings.neurips.cc/paper_files/paper/2025/file/0d904d300a105809a2114d727851e759-Paper-Conference.pdf}
}

@misc{openai2026models,
  author       = {{OpenAI}},
  title        = {{OpenAI API Models}},
  year         = {2026},
  howpublished = {\url{https://developers.openai.com/api/docs/models}},
  note         = {Accessed: 2026-05-27}
}

@misc{deepseekai2026v4release,
  author       = {{DeepSeek-AI}},
  title        = {{DeepSeek V4 Preview Release}},
  year         = {2026},
  howpublished = {\url{https://api-docs.deepseek.com/news/news260424}},
  note         = {Accessed: 2026-05-27}
}

\end{document}